\documentclass[prc,twocolumn,superscriptaddress,showpacs]{revtex4-1}
\usepackage{graphicx,amsmath,amssymb,bm,nicefrac,color}

\newcommand{\be}{\begin{equation}}	
\newcommand{\ee}{\end{equation}}
\newcommand{\vlowk}{V_{{\rm low}\,k}}
\newcommand{\lm}{\Lambda}
\newcommand{\fmi}{\, \text{fm}^{-1}}
\newcommand{\mev}{\, \text{MeV}}
\newcommand{\kev}{\, \text{keV}}

\newcommand{\hw}{\hbar\omega}
\newcommand{\sdfp}{s d f_{\nicefrac{7}{2}} p_{\nicefrac{3}{2}}}
\newcommand{\du}{d_{\nicefrac{5}{2}}}
\newcommand{\dl}{d_{\nicefrac{3}{2}}}
\newcommand{\s}{s_{\nicefrac{1}{2}}}
\newcommand{\f}{f_{\nicefrac{7}{2}}}
\newcommand{\p}{p_{\nicefrac{3}{2}}}

\begin{document}

\title{Chiral three-nucleon forces and bound excited states in 
neutron-rich oxygen isotopes }

\author{J.\ D.\ Holt}
\affiliation{Department of Physics and Astronomy, University of Tennessee, 
Knoxville, TN 37996, USA}
\affiliation{Physics Division, Oak Ridge National Laboratory, P.O. Box 2008,
Oak Ridge, TN 37831, USA}
\author{J.\ Men\'{e}ndez}
\affiliation{Institut f\"ur Kernphysik, Technische Universit\"at
Darmstadt, 64289 Darmstadt, Germany}
\affiliation{ExtreMe Matter Institute EMMI, GSI Helmholtzzentrum f\"ur
Schwerionenforschung GmbH, 64291 Darmstadt, Germany}
\author{A.\ Schwenk}
\affiliation{ExtreMe Matter Institute EMMI, GSI Helmholtzzentrum f\"ur
Schwerionenforschung GmbH, 64291 Darmstadt, Germany}
\affiliation{Institut f\"ur Kernphysik, Technische Universit\"at
Darmstadt, 64289 Darmstadt, Germany}

\begin{abstract}
We study the spectra of neutron-rich oxygen isotopes based on chiral
two- and three-nucleon interactions. First, we benchmark our
many-body approach by comparing ground-state energies to
coupled-cluster results for the same two-nucleon interaction, with
overall good agreement. We then calculate bound excited states in
$^{21,22,23}$O, focusing on the role of three-nucleon forces, in the
standard $sd$ shell and an extended $\sdfp$ valence space. Chiral
three-nucleon forces provide important one- and two-body
contributions between valence neutrons. We find that both these
contributions and an extended valence space are necessary to
reproduce key signatures of novel shell evolution, such as the
$N=14$ magic number and the low-lying states in $^{21}$O and
$^{23}$O, which are too compressed with two-nucleon interactions
only. For the extended space calculations, this presents first
work based on nuclear forces without adjustments. Future work is 
needed and open questions are discussed.
\end{abstract}


\maketitle

{\it Introduction.--} The oxygen isotopes provide an exciting
laboratory, both experimentally and theoretically, to study the
structure of extreme neutron-rich nuclei towards and beyond the
neutron dripline at $^{24}$O~\cite{Thoennessen}. Since $^{24}$O is
anomalously close to the valley of stability, all bound oxygen
isotopes have been explored experimentally~\cite{24O,Stanoiu,%
22Ospins,CortinaGil,ElekesSchiller,Hoffman,Fernandez}. The
neutron-rich isotopes exhibit surprising new features, such as the
closed-shell properties of $^{22}$O and $^{24}$O, and the
corresponding appearance of new magic numbers $N=14$ and 
$N=16$~\cite{magic,Oxygen}. The spectra with these
fingerprints provide a challenge for nuclear theory, and a unique test
of nuclear forces and many-body methods for neutron-rich nuclei.

While the chain of oxygen isotopes can be explored by different
theoretical methods~\cite{Oxygen,CCoxygen,CCoxygen2,continuum,CCSF}, recent
studies have focused on ground-state properties relevant for
understanding the location of the neutron dripline. In theories based
on two-nucleon (NN) forces only, the interactions were found to be too
attractive and to predict bound oxygen isotopes to
$^{28}$O~\cite{Oxygen}. Coupled-cluster (CC) calculations obtained
bound isotopes to $^{25}$O with chiral NN forces only, but found the
location of the dripline to depend on the resolution scale, implying
that higher-body forces need to be considered~\cite{CCoxygen}. The
first work including three-nucleon (3N) forces demonstrated that they
play a decisive role for the oxygen anomaly and can explain why
$^{24}$O is the heaviest bound oxygen isotope~\cite{Oxygen}.  Subsequent
CC calculations implementing adjusted 3N  
effects and coupling to the continuum have found similar 
results~\cite{CCoxygen2}.

As a result,
phenomenological shell-model interactions~\cite{USD,SDPFM,Nowacki}
have been developed that are adjusted to correct for these
deficiencies. As suggested by Zuker~\cite{Zuker}, we have shown that
these phenomenological adjustments can be largely traced to neglected
3N forces, both for the oxygen ground-state
energies~\cite{Oxygen,mono} and in calcium where 3N forces are key for
the $N=28$ magic number~\cite{Calcium}. 

\begin{figure*}[t]
\begin{center}
\includegraphics[scale=0.66,clip=]{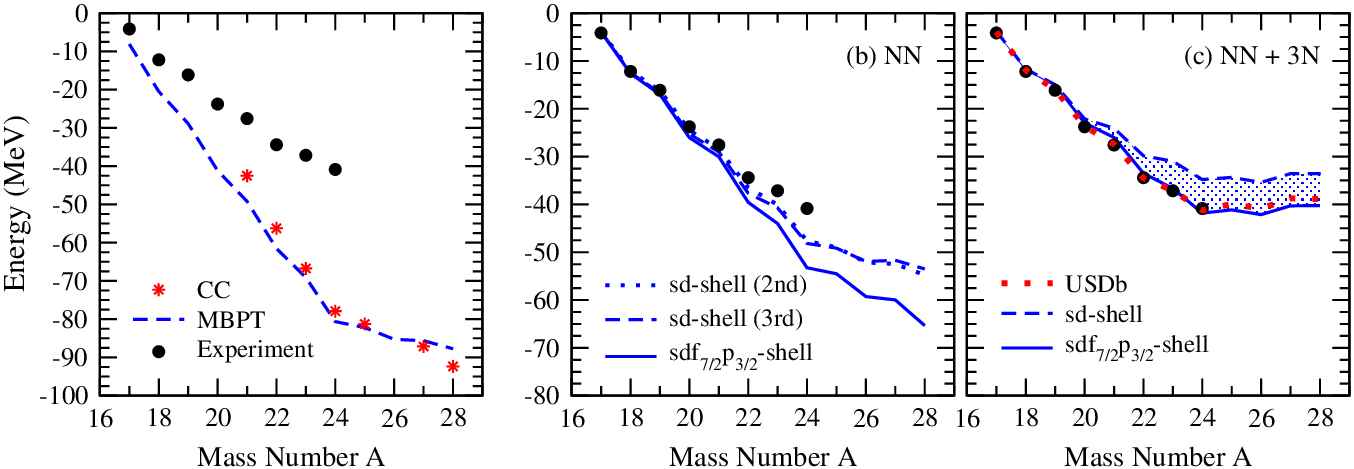}
\end{center}
\caption{Ground-state energies of neutron-rich oxygen isotopes
relative to $^{16}$O, with experimental energies from the AME 2003
atomic mass evaluation~\cite{AME2003}. Panel~(a) compares results
calculated in MBPT to those from CC theory based on the
same low-momentum NN interactions. The MBPT results are based on the
SPEs obtained in CC theory, for details see
text. Panel~(b) shows the NN-only MBPT results at 2nd and 3rd order
in the $sd$ shell, and 3rd order in the $\sdfp$ shell, using
empirical SPEs. In panel~(c), we present the energies
obtained at 3rd order from NN and 3N forces, with calculated
MBPT SPEs of Table~\ref{spetab} that also include 3N contributions,
in comparison to the phenomenological USDb model. The band indicates
the energy difference between the two valence spaces considered.\label{gs}}
\end{figure*}

Three-nucleon forces arise because nucleons are finite-mass composite
particles that can be excited by interacting with other particles. The
pioneering work of Fujita and Miyazawa~\cite{FM} established the
dominance of the single-Delta-excitation mechanism in the long-range
part of 3N forces, where one nucleon virtually excites a second
nucleon to the $\Delta(1232 \mev)$ resonance that is then de-excited
by interacting with a third nucleon. Additional long- and
shorter-range 3N interactions are included naturally in chiral
effective field theory (EFT)~\cite{chiral}, which provides a
systematic expansion for nuclear forces. Similar to the oxygen and
calcium isotopes, chiral 3N forces give repulsive contributions to the
neutron and nuclear matter energy~\cite{nucmatt,Jeremy,Kai} and to
pairing gaps in nuclei~\cite{pairing}.

In this paper, we present the first study of the impact of chiral 3N
forces on bound excited states in neutron-rich oxygen isotopes. We
first discuss the forces and many-body approach used, and benchmark
the ground-state energies with CC theory based on the same NN
forces. We then calculate the spectra of $^{21,22,23}$O in the
standard $sd$ shell and an extended $\sdfp$ valence space, focusing on
the role of 3N forces. Our results show that one- and two-body
contributions from chiral 3N forces to valence neutrons, included to
third order in our many-body formalism, as well as a
valence space larger than the standard $sd$ shell are important to
understand the experimental excited state structures.
Extended space calculations present an additional challenge,
because potential center-of-mass (cm) admixture must be addressed;
we discuss several approaches to clarify this issue.

{\it Interactions among valence neutrons.--} We calculate the
interactions among valence neutrons based on chiral NN and 3N forces
following Refs.~\cite{Calcium,Gallant}. At the NN level, we first
perform a renormalization-group (RG) evolution~\cite{Vlowk} of the
N$^3$LO NN potential of Ref.~\cite{N3LO} to a lower resolution scale
$\lm = 2.0 \fmi$. The resulting low-momentum interaction $\vlowk$ is
used to calculate the two-body interactions among valence neutrons to
third order in many-body perturbation theory (MBPT), following the
formalism of Ref.~\cite{LNP+Gmatrix}. We use a harmonic oscillator
basis of 13 major shells with $\hbar \omega = 13.53 \mev$, appropriate
for the oxygen isotopes.  In the MBPT calculations including 3N
forces, we also scale the resulting valence-shell interaction and
bound single-particle energies (SPEs) by $\hbar\omega \sim A^{-1/3}$.
The MBPT results are converged in terms of intermediate-state
excitations. In addition to the standard $sd$ shell, we also consider
an extended $\sdfp$ valence space, where the largest changes are due
to including the lower-lying $\f$ orbital.

To study the validity of the MBPT approach, we have carried out CC
calculations for the ground-state energies of the oxygen isotopes,
using the same $\vlowk$ interaction and the same basis space of 13
major shells with $\hbar \omega=14 \mev$. The CC energies are well
converged, with a flat $\hw$ dependence of $\sim 200 \kev$ change from
$\hw=14 - 20 \mev$. The results are shown in Fig.~\ref{gs}~(a)
relative to the ground-state energy of $^{16}$O. The closed
$j$-subshell systems, $^{16,22,24,28}$O, are calculated at the
$\Lambda$-CCSD(T) level~\cite{CClong} (with small differences from
CCSD $\sim 1 \mev$). The $A \pm 1$ systems, $^{17,21,23,25,27}$O, are
obtained with the CC particle-attached/removed equations of motion
method at the singles and doubles level
(PA/PR-EOM-CCSD)~\cite{CClong}.

For the CC comparison, we perform the MBPT calculations in the $sd$
shell, where for consistency the SPEs are taken as the PA-EOM-CCSD
($\du$, $\s$, $\dl$) energies in $^{17}$O. The particle-attached $\f$
and $\p$ energies are not single particle in character (at this NN
level), so the MBPT $sd$-shell comparison provides the cleanest
benchmark. In Fig.~\ref{gs}~(a), we find that the MBPT energies agree
within a few percent with CC theory (the agreement of excitation
energies is expected to be even better). This shows that, at this
level, MBPT can be comparable to CC theory for $\vlowk$ interactions,
but also highlights the important role of SPEs. Furthermore, comparing
with the results shown in Figs.~\ref{gs}~(b) or~(c), we see that the
addition of 3N forces or going to an extended valence space has a
greater impact on the ground-state energies than the uncertainty in
MBPT in Fig.~\ref{gs}~(a).

At the 3N level, we include 3N forces among two valence neutrons and
one nucleon in the core. This corresponds to the normal-ordered
two-body part of 3N forces, which was found to dominate in CC
calculations~\cite{CC3N} over residual 3N forces (the latter
contributions should be weaker for normal Fermi
systems~\cite{Fermi}). We take into account chiral 3N forces at
N$^2$LO~\cite{chiral3N} that give rise to repulsive monopole
interactions between valence neutrons~\cite{Oxygen,Calcium}. At
N$^2$LO, this includes long-range two-pion-exchange parts $c_i$ 
(due to $\Delta$ and other excitations), plus shorter-range one-pion
exchange $c_D$ and 3N contact $c_E$ interactions, with the $c_D, c_E$
couplings fit to the $^3$H binding energy and the $^4$He radius for
$\Lambda = \Lambda_{\rm 3N} =2.0 \fmi$~\cite{3Nfit}. The fit to light
nuclei approximately includes effects of the RG evolution in the 3N
sector, up to higher-order many-body forces. For all results, these 3N forces,
within 5 major oscillator shells, are included fully to third order in MBPT.
Future work will be to study the consistent RG evolution of 3N forces
and to include N$^3$LO 3N and 4N forces~\cite{chiral}.

For the SPEs in $^{17}$O, we solve the Dyson equation, consistently
including one-body contributions to third order in MBPT in the same
space as the two-body interactions. While the convergence in terms of
intermediate-state excitations is slower for the one-body energies,
the results are still well converged in 13 major shells. In the Dyson
equation, we include chiral 3N forces between one valence neutron and
two core nucleons, which corresponds to the normal-ordered one-body
part of 3N forces. The 3N contributions range between $1.5-3.0 \mev$,
approximately an order of magnitude larger than typical normal-ordered
two-body matrix elements. This is consistent with the normal-ordering
hierarchy~\cite{CC3N,Fermi}. At each self-consistency iteration, the
unperturbed harmonic oscillator spectrum is updated such that the
energy of the degenerate valence-space orbitals is set to the
centroid, $\sum_j (2j+1) \varepsilon_j/(2j+1)$, of the calculated SPEs
of the previous iteration.

Our results for the SPEs in the $sd$ and $\sdfp$ valence spaces are
given in Table~\ref{spetab}, in comparison with empirical SPEs taken
from the USDb~\cite{USD} and SDPF-M~\cite{SDPFM} models. With 3N
forces included, the resulting SPEs are similar to the empirical
values, except that the $\p$ orbit is significantly higher.
Note that the $\f$ orbital is obtained only $2\mev$ above 
the $sd$ shell, allowing sizeable contributions from this orbital.
It will be important to investigate this issue in improved 
many-body calculations.
Because the higher-lying orbitals are unbound, a necessary future improvement
will be to represent these correctly as continuum states.  Work in
this direction is under development, but CC calculations in this
region have found that continuum coupling effectively lowers energies
of unbound states only by $\sim300\kev$~\cite{CCoxygen2}.

\begin{table}
\begin{center}
\caption{Empirical and calculated (MBPT) SPEs in MeV.\label{spetab}}
\begin{tabular*}{0.48\textwidth}{c|cc|cc}
\hline\hline
\, orbital \, & \multicolumn{2}{|c}{$sd$ shell} &  
\multicolumn{2}{|c}{$\sdfp$ shell} \\
& \, USDb~\cite{USD} \, & \, MBPT \, 
& \, SDPF-M~\cite{SDPFM} \, & \, MBPT \, \\ \hline
$\du$ & $-3.93$ & $-3.78$ & $-3.95$ & $-3.46$ \\ 
$\s$  & $-3.21$ & $-2.42$ & $-3.16$ & $-2.20$ \\ 
$\dl$ &  $\,\,\,\,\,2.11$ & $\,\,\,\,\,1.45$ & $\,\,\,\,\,1.65$ & 
$\,\,\,\,\,1.92$ \\ 
$\f$ & $-$ & $-$ & $\,\,\,\,\,3.10$ & $\,\,\,\,\,3.71$ \\
$\p$ & $-$ & $-$ & $\,\,\,\,\,3.10$ & $\,\,\,\,\,7.72$ \\[0.5mm] 
\hline\hline
\end{tabular*}
\end{center}
\end{table}

Following our MBPT strategy, our goal is to work in the largest
possible valence space for the diagonalization, which leads to valence
spaces involving more than one major harmonic oscillator shell. In
such spaces, however, the cm motion does not
factorize in general. As in Refs.~\cite{ZukerCM,Ca40CM}, we
investigate the possible admixture by adding a cm Hamiltonian, $\beta
H_{cm}$, with $\beta=0.5$, to our original Hamiltonian. While the
qualitative picture of our results remains, there is some additional
compression in the studied spectra as a result
(e.g., the $2^+_1$ states in $^{22}$O and $^{24}$O are 
lowered by $0.9 \mev$).
This is not unexpected, since the non-zero cm two-body matrix elements are also
relevant matrix elements of the MBPT calculation, and making a clear
separation between them is therefore difficult. We also find
non-negligible $\langle H_{cm} \rangle$ expectation values,
$9.1$ and $14.8 \mev$ for the ground states of $^{22}$O and $^{24}$O,
which points to possible cm admixture and/or non-neglible occupancies of the
$\f$ and $\p$ orbitals in the calculations of these neutron-rich nuclei.

However, as discussed in Ref.~\cite{cm} and shown in the context of CC
theory~\cite{CCCOM}, $\beta$-dependence and $\langle H_{cm} \rangle
\ne 0$ do not necessarily imply cm admixture.
Given the importance of undertanding this issue, 
work is in progress in several directions. First,
we will apply the singular-value decomposition
used in Ref.~\cite{CCCOM} to test cm factorization to
extended-space shell-model calculations. In addition, following the
MBPT strategy, we will carry out a nonperturbative
Okubo-Lee-Suzuki-Okamoto transformation~\cite{OLSO} into the standard
one-major-shell space, where the cm motion factorizes. This treats the
MBPT configurations perturbatively, but includes the lower-lying
extended-space orbitals nonperturbatively. Finally, we will also apply
the in-medium similarity renormalization group (IM-SRG)~\cite{IMSRG}
to extended valence spaces, where the cross-shell matrix elements can
be driven to small values under the IM-SRG evolution.

{\it Ground-state energies.--} Although the focus of this work is on
excited states, we first discuss results for the ground-state energies
in this framework. All calculations based on NN forces are performed
with empirical SPEs, and all calculations including 3N forces with the
consistently calculated MBPT SPEs of Table~\ref{spetab}.  We take into
account many-body correlations by diagonalization in the respective
valence spaces
using the shell model code Antoine~\cite{antoine,SMRMP}.
Figures~\ref{gs}~(b) and~(c) show the resulting
ground-state energies compared with experiment. With NN forces only,
all isotopes out to $^{28}$O remain bound because the two-body
interactions among valence neutrons are too
attractive~\cite{Oxygen}. Going to the extended $\sdfp$ space
provides additional attraction, mainly due to the correlations involving
the $\f$ orbital. To gauge the order-by-order convergence of
MBPT, we also show in Fig.~\ref{gs}~(b) the ground-state energies to
2nd order. The small difference from 2nd to 3rd order is consistent
with the CC agreement and shows that the MBPT calculations can be
converged to a good level.

When 3N forces are included in Fig.~\ref{gs}~(c), we see their clear
repulsive contribution to the ground-state energies, relative to
$^{16}$O~\cite{Oxygen}. In both valence spaces, there is significantly
less binding compared to the NN-only calculations.
The calculated energies in the $sd$ shell are underbound with respect
to experiment, while in the extended $\sdfp$ space, we find very good
agreement. Finally, we note that in the $\sdfp$
space the ground- and excited-state wavefunctions contain an admixture
of a $2p2h$ intruder configuration. In the NN+3N calculations, the 
dominant components of the ground-state wavefunctions, $\du$, $\f$, and 
$\s$, respectively, change from 3.7, 0.9, 0.2;
4.2, 1.3, 0.3; 4.4, 1.5, 0.9; and 4.4, 1.6, 1.7 from $^{21}$O through 
$^{24}$O, showing the expected $\s$ filling. In contrast, the 
corresponding NN-only ground-state wavefunction 
occupancies of $^{22}$O, 3.6, 0.9, 1.2, reveal a significant $\s$ 
component not present in the NN+3N calculations.

\begin{figure}
\begin{center}
\includegraphics[scale=0.365,clip=]{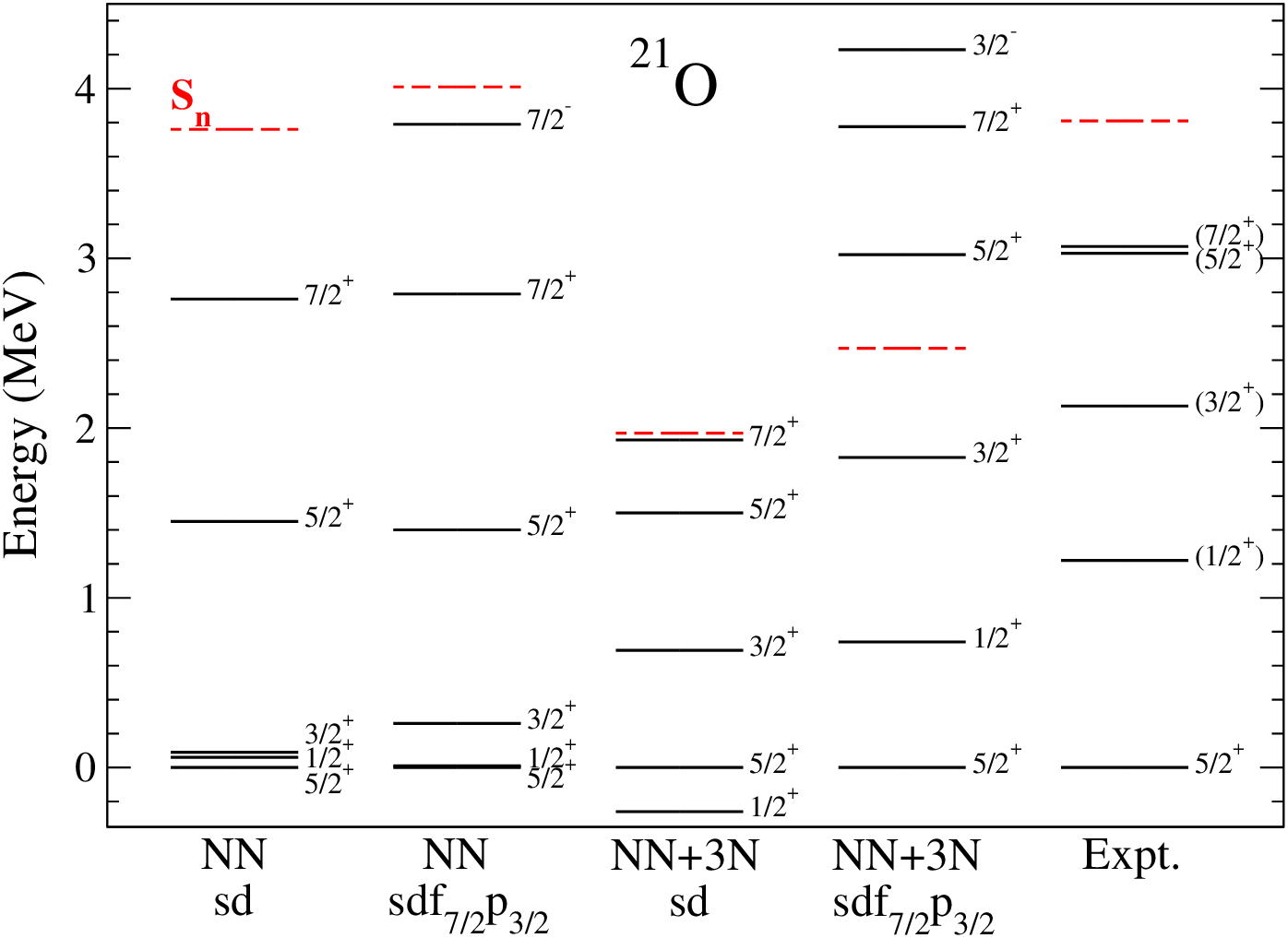}
\end{center}
\caption{Excitation energies of bound excited states in $^{21}$O
compared with experiment~\cite{Stanoiu,Fernandez}. The NN-only results are
calculated in the $sd$ and $\sdfp$ shells with empirical SPEs. The
NN+3N energies are obtained in the same spaces, but with calculated
MBPT SPEs, including 3N contributions to valence two-body
interactions and SPEs. The dashed lines give the
one-neutron separation energy $S_n$.\label{21O}}
\end{figure}

{\it Spectra.--} Next, we study the properties of bound excited states
in the neutron-rich oxygen isotopes $^{21,22,23}$O, which have been
the subject of recent experimental
investigations~\cite{Stanoiu,22Ospins,ElekesSchiller,Fernandez}. 
While we will not explicitly discuss spectra of the lighter isotopes,
we note that in the  NN+3N extended-space calculation, the
first excited $2^+$  ($4^+$) energies in $^{18}$O and  $^{20}$O 
are $2.25\mev$ ($3.90\mev$) and $2.14\mev$ ($4.99\mev$), respectively, 
compared with experimental values $1.98\mev$ ($3.55\mev$) and 
$1.67\mev$ ($3.57\mev$).  Furthermore, the calculated $E(4^+)/E(2^+)$ 
ratios of 1.79 and 2.33 for $^{18}$O and $^{20}$O compare well with 
the experimental values of 1.73 and 2.14. For each nucleus we 
discuss and compare with all measured states below the one-neutron 
separation energy $S_n$, including for reference the lowest-lying 
negative-parity state in the odd-mass isotopes. The coupling to the scattering
continuum is important for unbound states and as the energies
of excited states approach $S_n$~\cite{CCoxygen2,continuum,open}. 
Here, we focus on 3N force contributions and note possible 
continuum effects for the higher-lying states.

{\it $^{21}$O.--} The spectrum of $^{21}$O in Fig.~\ref{21O} shows
clearly the deficiencies of calculations based on NN forces.  In a
basic $sd$-shell picture, the ground state is interpreted as a
one-hole $\frac{5}{2}^+$ state below the $^{22}$O closure, while the
excited states result from a one-particle excitation to the $\s$
orbital coupled to two holes in $\du$ with angular momenta $0^+$,
$2^+$, and $4^+$ (a $\frac{9}{2}^+$ state is observed above the
neutron-decay threshold). These states therefore probe the $\du-\s$
gap. With NN forces only, the calculated spectrum is in very poor
agreement with experiment, being too compressed for the low-lying
states and too separated for the $\frac{5}{2}^+$ and $\frac{7}{2}^+$
states. Expanding the valence space to the $\sdfp$ shell provides only
slight improvement for the $\frac{3}{2}^+$ state, but otherwise has
little effect.

\begin{figure}
\begin{center}
\includegraphics[scale=0.365,clip=]{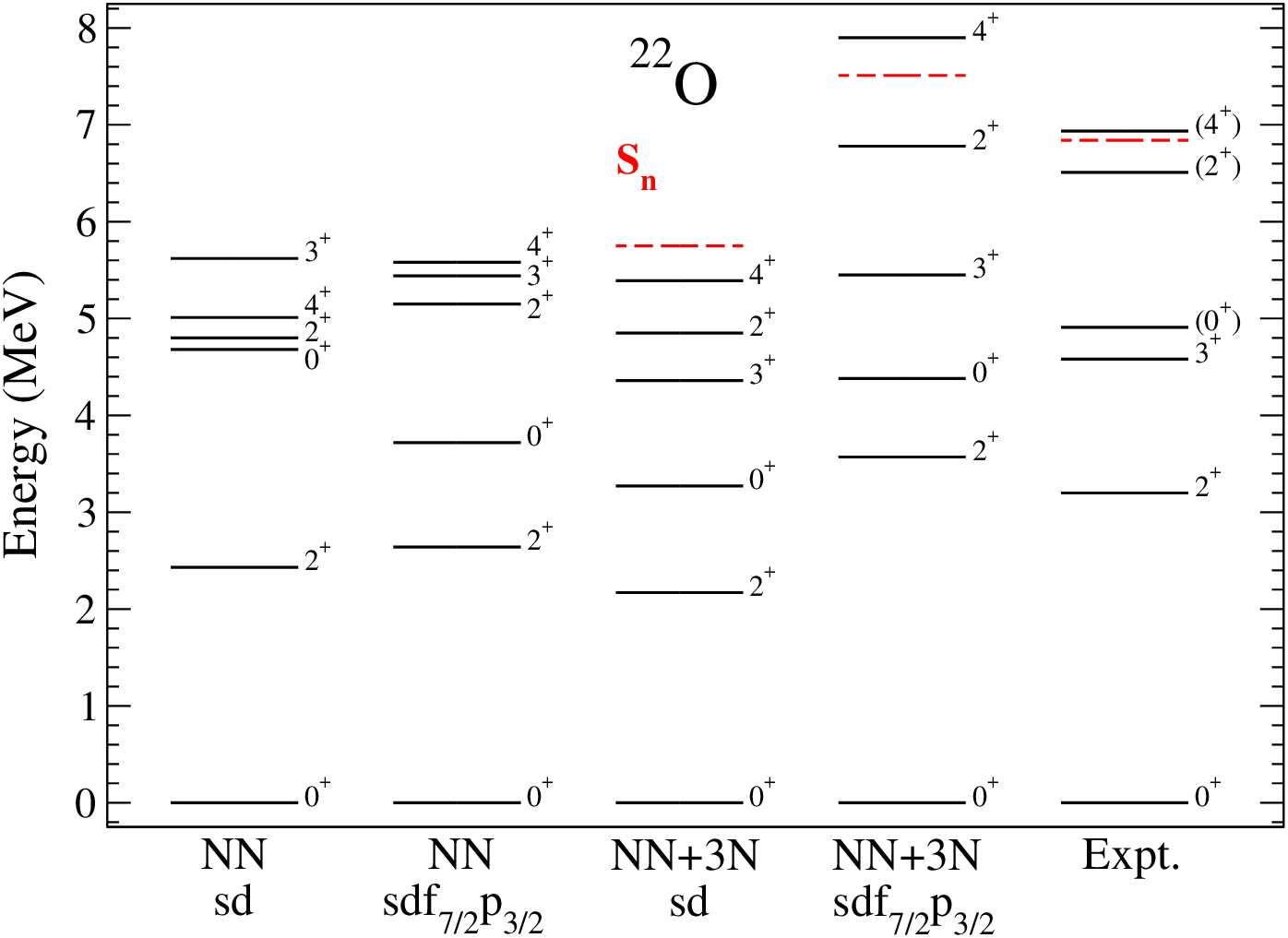}
\end{center}
\caption{Excitation energies of bound excited states in $^{22}$O
compared with experiment~\cite{Stanoiu,22Ospins} (labels are as in 
Fig.~\ref{21O}).\label{22O}}
\end{figure}

With the inclusion of 3N force contributions, the low-lying states
become more spread out, but with an incorrect $\frac{1}{2}^+$ ground state. 
For the $sd$ shell, $S_n$ is well
below the experimental value, which reflects the underbound
ground-state energies. Our results in the extended $\sdfp$ space with
3N forces show significant improvements, although the $\frac{1}{2}^+$
and $\frac{3}{2}^+$ states remain somewhat below the experimental
values.  Furthermore, due to their presence above the continuum
threshold, the $\frac{5}{2}^+_2$ and $\frac{7}{2}^+$ energies may 
be reduced, bringing them closer to experiment.  

{\it $^{22}$O.--} The first oxygen isotope exhibiting closed shell
properties for a non-standard magic number, $^{22}$O, has its first
$2^+$ state at almost twice the energy as those in $^{18}$O and
$^{20}$O. In contrast to $^{24}$O, whose closed-shell nature can be
qualitatively well described in NN-only calculations due to the large
separation between the $\dl$ and $\s$ SPEs (see, e.g.,
Ref.~\cite{CCSF}), the spectrum of $^{22}$O is not well reproduced
with NN forces, as shown in Fig.~\ref{22O}. In the $sd$-shell,
the ground state is predominantly a filled $\du$ subshell, while the
first excited $2^+$ and $3^+$ states result from a one-particle $\s$
excitation coupled to a $\du$ hole. The higher-lying $0^+$, $2^+$,
$4^+$ triplet is formed by two excited $\s$ particles coupling to two
$\du$ holes.

\begin{figure}
\begin{center}
\includegraphics[scale=0.365,clip=]{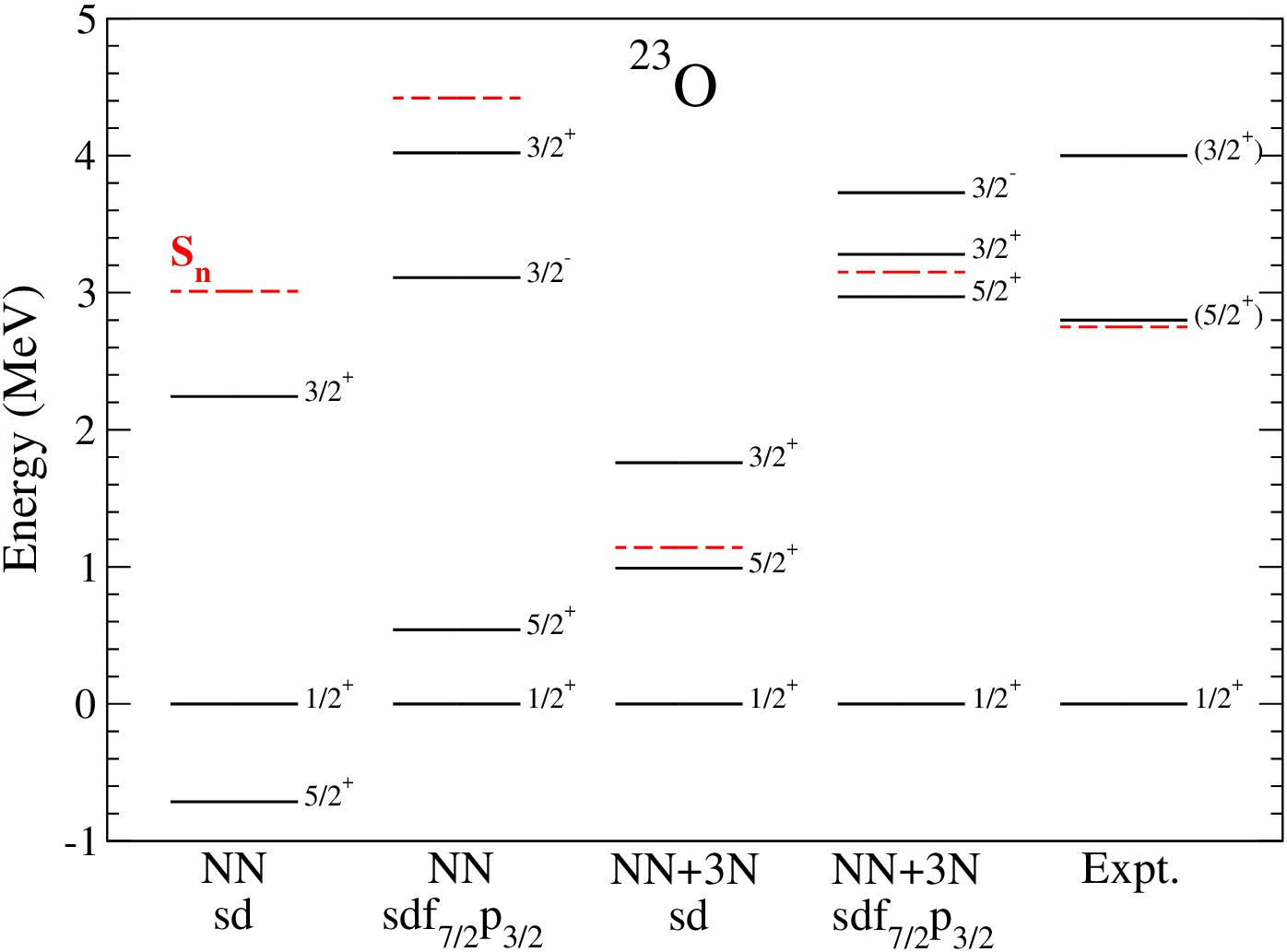}
\end{center}
\caption{Excitation energies of bound excited states in $^{23}$O
compared with experiment~\cite{ElekesSchiller} (labels are as
in Fig.~\ref{21O}).\label{23O}}
\end{figure}

With NN forces in the $sd$ shell, the first $2^+$ state is below
experiment, the spacing to the $3^+$ state too broad, and the $0^+$,
$2^+$, and $4^+$ states are too compressed. Extending the space
provides only a slight improvement compared to the experimental
spectrum.  The addition of 3N forces in the $sd$ shell reduces the
first $2^+$ energy but otherwise improves the level spacings.

It is only when we move to the $\sdfp$ space with 3N forces included
that we see significant improvement in the spectrum, with the $2^+_1$
energy now being approximately twice the $2^+_1$ energies of $^{18}$O
and $^{20}$O.  Although the $2^+-3^+$ splitting is reduced, it is
still somewhat large and results in an inversion of the $3^+$ and
$0^+$ states compared with experiment, a feature also seen in all
phenomenological USD models. Since the $4^+$ energy lies above $S_n$,
some lowering can be expected from continuum effects. Similar to
$N=28$ in calcium~\cite{Calcium}, we find that both 3N forces and an
extended valence space are important for reproducing properties of the
$N=14$ magic number in oxygen.

{\it $^{23}$O.--} Occupying a position of particular interest between
two doubly-magic isotopes, the spectrum of $^{23}$O provides a unique
test for theory, as it should simultaneously reflect the features of
both. The $sd$-shell $^{23}$O ground state is dominated by one particle in the
$\s$ orbit, while the two lowest excited states are expected to be a
single particle $\frac{5}{2}^+$ one-hole excitation, indicative of the
strength of the $^{22}$O shell closure, and a higher-lying single
particle $\frac{3}{2}^+$ one-particle excitation, reflecting the
strength of the $^{24}$O shell closure. Recent experimental
measurements of the lowest-lying states in $^{23}$O shown in
Fig.~\ref{23O} confirm this picture, although no bound excited states
are found~\cite{ElekesSchiller}. The $\frac{5}{2}^+$ state lies just
above the neutron decay threshold, while the $\frac{3}{2}^+$ state
resides well in the continuum, and a complete description cannot 
ignore these effects.

{}From Fig.~\ref{23O} we see that the $\frac{5}{2}^+$ state is bound in
all calculations. With NN forces in the $sd$ shell, the
$\frac{5}{2}^+$ state even lies below the $\frac{1}{2}^+$, and the
$\frac{3}{2}^+$ is bound, consistent with the expectation of a bound
$\dl$ orbital at $^{24}$O without 3N forces~\cite{Oxygen}.  Extending
the space to the $\sdfp$ orbitals is a clear improvement, and the
$\frac{5}{2}^+$ is now $0.54 \mev$ above the $\frac{1}{2}^+$ ground
state. This is similar to the $0.35 \mev$ excitation energy obtained
in CC theory based on a N$^3$LO NN potential only~\cite{CCSF}.

The repulsive 3N contributions improve the spectrum considerably. We
again find the best agreement with experiment in the extended $\sdfp$
space, where the enhanced $\du-\f$ correlations, important for the
high $2^+$ in $^{22}$O, bring the $\frac{5}{2}^+$ energy very close
to the experimental value. This shows that 3N forces are essential for
the $\frac{5}{2}^+$ state, increasing the excitation energy by $\sim 2
\mev$ compared to the NN-only result. The $\frac{3}{2}^+$ state, however, 
is significantly lower than experiment lying just beyond the neutron-decay
threshold. When continuum coupling is included, this will effectively
lower the $\dl$ orbit~\cite{continuum}, pointing to the need
to explore uncertainties as well as further improvements in our 
theoretical approach.

{\it Summary.--} We have presented the first study of the impact of
chiral 3N forces on the spectra of neutron-rich oxygen isotopes.
An important next step is to study the theoretical uncertainties
of the results, due to the truncation in the chiral EFT expansion,
to uncerainties in the low-energy couplings in 3N forces, and due to the
many-body calculation. Based
on the MBPT approach and chiral EFT interactions, we find that the
traditional $sd$ shell is insufficient, even with 3N forces, to
account for key experimental features in the spectra, in particular
the $N=14$ magic number. Going to an extended $\sdfp$ valence space,
and including NN and 3N contributions to third order in MBPT
consistently to the two-body interactions and the one-body SPEs, our
results are in good overall agreement with experiment.   For the
extended space calculations, this presents first work based on
nuclear forces without adjustments. Future work is needed, in
particular to understand and quantify the potential cm admixture. We are
currently investigating this issue with different approaches.
It will also be interesting to study how the physics of the extended
valence space may be renormalized into phenomenological $sd$-shell
models. In addition, extending our framework to include continuum
effects for weakly-bound or unbound states, and for nuclei beyond the
driplines, as well as developing nonperturbative approaches to valence
shell interactions (see, e.g., Refs.~\cite{NCSMcore,IMSRG,2pattached})
are important future steps.

We thank G.\ Hagen, T.\ Otsuka and T.\ Suzuki for very
useful discussions, and the TU Darmstadt for hospitality. This work
was supported in part by US DOE Grant DE-FC02-07ER41457 (UNEDF
SciDAC Collaboration), DE-FG02-96ER40963 (UT) and DE-FG02-06ER41407
(JUSTIPEN), by the Helmholtz Alliance Program of the Helmholtz
Association, contract HA216/EMMI ``Extremes of Density and
Temperature: Cosmic Matter in the Laboratory'', and the DFG through
Grant SFB 634. Computations were performed with an allocation of
advanced computing resources on Kraken at the National Institute for
Computational Sciences.

\end{document}